\documentclass[12pt,english]{article}
\usepackage{mathptmx}
\usepackage[T1]{fontenc}
\usepackage[latin9]{inputenc}
\usepackage{geometry}
\usepackage{babel}
\usepackage{float}
\usepackage{units}
\usepackage{bm}
\usepackage{amsmath}
\usepackage{amssymb}
\usepackage{graphicx}
\usepackage{setspace}
\usepackage[authoryear]{natbib}
\usepackage[unicode=true,
 bookmarks=true,bookmarksnumbered=false,bookmarksopen=false,
 breaklinks=false,pdfborder={0 0 1},backref=false,colorlinks=false]
 {hyperref}

\makeatletter

\providecommand{\tabularnewline}{\\}

\usepackage{bm}
\let\newcommand=\providecommand

\usepackage[format=hang,font=small,labelfont=bf]{caption}

\makeatother

\begin{document}
\begin{center}
\textbf{\Large{}A different approach for choosing a threshold in peaks
over threshold}{\Large\par}
\par\end{center}

\begin{center}
Andr\'ehette Verster$^{1}$\footnote{Corresponding author. E-mail: verstera@ufs.ac.za}
and Lizanne Raubenheimer\emph{\small{}$^{2}$}\\
{\small{}$^{1}$Department of Mathematical Statistics and Actuarial
Science, University of the Free State, Bloemfontein, South Africa}\\
{\small{}$^{2}$School of Mathematical and Statistical Sciences, North-West
University, Potchefstroom, South Africa}{\small\par}
\par\end{center}
\begin{abstract}
In Extreme Value methodology the choice of threshold plays an important
role in efficient modelling of observations exceeding the threshold.
The threshold must be chosen high enough to ensure an unbiased extreme
value index but choosing the threshold too high results in uncontrolled
variances. This paper investigates a generalized model that can assist
in the choice of optimal threshold values in the $\gamma$ positive
domain. A Bayesian approach is considered by deriving a posterior
distribution for the unknown generalized parameter. Using the properties
of the posterior distribution allows for a method to choose an optimal
threshold without visual inspection.\emph{ }
\end{abstract}

\section{Introduction}

In Extreme Value methodology we assume that a random sample $X_{1},...,X_{n}$
from an unknown distribution $F$ belongs to the domain of attraction
of a generalized extreme value distribution. Let $X_{n,n}=\text{max}\left(X_{i}\text{'s}\right)$
and assume there exists sequences $a_{n}>0$ and $b_{n}$ such that
as $n\rightarrow\infty$, 
\begin{align}
P\left(\frac{X_{n,n}-b_{n}}{a_{n}}\right) & \rightarrow G_{\gamma}\left(x\right),\label{eq:1}
\end{align}
 where $\gamma$ is known as the extreme value index (EVI), Beirlant et al. (2004).
All extreme value distributions, $G_{\gamma}\left(x\right)=\text{exp}\left[-\left(1+\gamma x\right)^{-\nicefrac{1}{\gamma}}\right]$
for $1+\gamma x>0$, $x\in\mathbb{R}$, can occur as limits in (\ref{eq:1}).
It is well known that 
\begin{align}
P\left(X-u>y\left|X>u\right.\right) & =\frac{\overline{F}\left(u+y\right)}{\overline{F}\left(u\right)}\rightarrow\overline{H}_{\gamma}\left(y\right)=\left(1+\gamma y\right)^{-\nicefrac{1}{\gamma}},\;\;x>u,\;\;\gamma\in\mathbb{R}.\label{eq:2}
\end{align}

See for example Beirlant et al. (2004) and de Haan \& Ferreira (2005) for further details. From (\ref{eq:2}) it states that the excesses of the conditional
distribution of $X-u$, given $X>u$ can be modelled through a peaks
over threshold (POT) model, the generalized Pareto distribution (GPD)
given in (\ref{eq:2}) as $H_{\gamma}$. Very often in literature
the focus is only on the Pareto type case where $\gamma$ in (\ref{eq:1})
is positive. The survival function of the Pareto type distribution
is given as 
\begin{align}
\overline{F}\left(x\right) & =x^{-\gamma}l\left(x\right),\;\;x>1,\label{eq:3}
\end{align}
 where $l\left(x\right)$ is known as the slowly varying function
that satisfies $\frac{l\left(xu\right)}{l\left(u\right)}\rightarrow1$
as $u\rightarrow\infty$ (Beirlant et al.,
2004), and $\nicefrac{1}{\gamma}$
is the EVI. Pareto-type distributions follows a simpler POT limit
as $u\rightarrow\infty$: 
\begin{align}
P\left(\frac{X}{u}>y\left|X>u\right.\right) & \rightarrow\overline{H}_{\gamma}\left(y\right)=y^{-\gamma},\;\;y>1.\label{eq:4}
\end{align}
 For the remainder of the paper (\ref{eq:4}) will be referred to
as the survival function of the Strict Pareto distribution. 

The limit results in (\ref{eq:2}) and (\ref{eq:4}) entails that
the threshold $u$ should be chosen as large as possible such that
the bias of $\gamma$ is not too large. In contrast to that, the threshold
$u$ should be chosen as small as possible such that the estimation
variation is limited. This is a complicated trade off, which makes
the threshold choice difficult in practice. Often the threshold is
chosen as one of the high data points in an ordered sample $X_{1,n}\leq X_{2,n}\leq\cdots\leq X_{n,n}$.

Several procedures have been proposed for reducing the bias and limiting
the variance, where a second order (sometimes third order) of (\ref{eq:4})
is assumed, ssee for example Feuerverger \& Hall (1999), Gomes et al.
(2000), Beirlant et al. (1999, 2009, 2019) and Caeiro \& Gomes (2011) to name a few. This paper slightly touches on
the bias reduction and variance stability of the EVI estimate but
is not an extensive study on this topic. The paper\textquoteright s
focus is on the investigation of a practical and rather simple method
to choose an optimal threshold value if the data follows a POT distribution
(with $\gamma>0$). Often in practice the EVI in (\ref{eq:4}) is
estimated, through various methods, such as maximum likelihood (ML)
or Method of moments at different threshold values. These estimates
are then plotted against the various threshold values, and the graph
is visually inspected to find a threshold range where the EVI estimate
seems to be stable. Although it is a popular method in practice it
is not always a clear picture with a lot of volatility. It is sometimes
difficult to see a stable area. The paper is arranged as follows:
In Section \ref{sec:Topp-Leone-Pareto-distribution} a Topp-Leone
Pareto distribution is considered where a generalization parameter
is introduced to generalize the strict Pareto distribution. This additional
parameter gives valuable insight to the choice of threshold. In Section
(\ref{sec:Bayesian-Analysis}), a Bayesian, rather than a classical
approach is considered in the parameter estimation and modelling process.
A small simulation study is conducted in Section \ref{sec:Simulation-study}
to investigate the effectiveness of the generalized model. The generalized
model is applied to a real data set in Section \ref{sec:Real-data-example}.
Section \ref{sec:Procedure-to-choose} explains the method we propose
for choosing an optimum threshold. 

\section{Topp-Leone Pareto distribution \label{sec:Topp-Leone-Pareto-distribution}}

The Topp \& Leone (1955) distribution was generalized by Rezaei et al. (2017) where $0\leq x\leq1$ is replaced by die CDF of any baseline distribution.
The CDF and probability density of the Topp-Leone generated family
of distributions are given by
\begin{align}
F\left(x;\alpha,\boldsymbol{\bm{\theta}}\right) & =\left\{ G\left(x;\boldsymbol{\bm{\theta}}\right)\left[2-G\left(x;\boldsymbol{\bm{\theta}}\right)\right]\right\} ^{\alpha}=\left\{ 1-\left[1-G\left(x;\boldsymbol{\bm{\theta}}\right)\right]^{2}\right\} ^{\alpha}\label{eq:5}
\end{align}
and
\begin{align}
f\left(x;\alpha,\boldsymbol{\bm{\theta}}\right) & =2\alpha g\left(x;\boldsymbol{\bm{\theta}}\right)\left[1-G\left(x;\boldsymbol{\bm{\theta}}\right)\right]\left\{ 1-\left[1-G\left(x;\boldsymbol{\bm{\theta}}\right)\right]^{2}\right\} ^{\alpha-1}\label{eq:6}
\end{align}
where $g\left(x;\boldsymbol{\bm{\theta}}\right)$ and $G\left(x;\boldsymbol{\bm{\theta}}\right)$
denote the probability density and CDF of the baseline distribution
with parameter set $\boldsymbol{\bm{\theta}}$, Razaei et al. (2017).
In this study we consider the baseline distribution as the Strict
Pareto distribution with probability density and CDF given respectively
by
\begin{align}
g\left(y;\gamma\right) & =\gamma\left(y\right)^{-\gamma-1},\;y=\frac{x}{u}>1,\;\gamma>0\label{eq:7}
\end{align}
and
\begin{align}
G\left(y;\gamma\right) & =1-y^{-\gamma},\;y=\frac{x}{u}>1,\;\gamma>0,\label{eq:8}
\end{align}
where $\nicefrac{1}{\gamma}$ denotes the EVI . Substituting (\ref{eq:7})
and (\ref{eq:8}) into (\ref{eq:5}) and (\ref{eq:6}) yields the
Topp-Leone Pareto (TLPa) distribution with PDF and CDF given respectively
as
\begin{align}
f\left(y;\alpha,\gamma\right) & =2\alpha\gamma y^{-2\gamma-1}\left(1-y^{-2\gamma}\right)^{\alpha-1},\label{eq:10}
\end{align}
and
\begin{align}
F\left(y;\alpha,\gamma\right) & =\left(1-y^{-2\gamma}\right)^{\alpha}\label{eq:9}
\end{align}
where $y>1$, $\gamma>0$ and $\alpha>0$.

The survival function of the TLPa can be expanded through a Binomial
expansion such that
\begin{align}
1-F\left(y;\alpha,\gamma\right) & =y^{-2\gamma}\left[\alpha-\frac{\alpha\left(\alpha-1\right)}{2}y^{-2\gamma}+o\left(y^{-2\gamma}\right)\right]\label{eq:11}
\end{align}
where $\alpha-\frac{\alpha\left(\alpha-1\right)}{2}y^{-2\gamma}+o\left(y^{-2\gamma}\right)$
is the slowly varying function and $\nicefrac{1}{2\gamma}$ is the
EVI. If $\alpha=1$ (\ref{eq:9}) and (\ref{eq:10}) become the Strict
Pareto probability and CDF with an EVI of $\nicefrac{1}{2\gamma}$.
In this study we are focusing our analysis on the assumption that
as $n\rightarrow\infty$ the relative excesses $\left(\nicefrac{X}{u}\right)$
will follow a Strict Pareto distribution (this assumption is well-known
in literature) or equivalently a TLPa with $\alpha=1.$ 

\section{Bayesian Analysis\label{sec:Bayesian-Analysis}}

It can easily be seen that when a Jeffreys prior, $p\left(\gamma\right)\propto\nicefrac{1}{\gamma}$,
is assumed for $\gamma$ in the Strict Pareto case, the posterior
distribution, given the data and the threshold, is given as:
\begin{align}
p\left(\gamma\left|\boldsymbol{\bm{y}},u\right.\right) & \propto\left[\prod_{i=1}^{n}\gamma\left(y_{i}\right)^{-\gamma-1}\right]\gamma^{-1}=\gamma^{n}\left(\prod_{i=1}^{n}\gamma e^{-\gamma\text{log}y_{i}}\right)\gamma^{-1}\nonumber \\
 & \propto\gamma^{n-1}e^{-\gamma\left(\sum\limits _{i=1}^{n}\text{log}y_{i}\right)}.\label{eq:12}
\end{align}
From (\ref{eq:12}) it is clear that the posterior is a gamma distribution
with parameters $n$ and $\sum_{i=1}^{n}\text{log}y_{i}$, where $n$
denotes the number of observations above the threshold. An estimate
of $\gamma$ can be the mean of the above gamma distribution. The
EVI of the Strict Pareto will then follow an inverse gamma distribution,
$\text{EVI}_{\text{SP}}\sim\text{InvGamma}\left(n,\left(\sum_{i=1}^{n}\text{log}y_{i}\right)^{-1}\right)$,
where SP represents the Strict Pareto case and the second parameter
is the rate parameter.

In the case of the TLPa the joint posterior is derived by assuming
an independent Jeffreys prior, $p\left(\gamma,\alpha\right)\propto\nicefrac{1}{\gamma\alpha}$
and the likelihood is given as
\begin{align}
p\left(\boldsymbol{\bm{y}}\left|\gamma,\alpha\right.\right) & \propto\alpha^{n}\gamma^{n}\prod_{i=1}^{n}y_{i}^{-2\gamma-1}\left(1-y_{i}^{-2\gamma}\right)^{\alpha-1}.\label{eq:13}
\end{align}
The joint posterior will be 
\begin{align}
p\left(\gamma,\alpha\left|\boldsymbol{\bm{y}}\right.\right) & \propto\alpha^{n-1}\gamma^{n-1}e^{-\left(2\gamma+1\right)\sum_{i=1}^{n}\text{log}\left(y_{i}\right)}e^{\left(\alpha-1\right)\sum_{i=1}^{n}\text{log}\left(1-y_{i}^{-2\gamma}\right)}.\label{eq:14}
\end{align}
The conditional posteriors can be approximated as
\begin{align}
p\left(\alpha\left|\gamma,\boldsymbol{\bm{y}}\right.\right) & \propto\alpha^{n-1}e^{-\alpha\left[-\sum_{i=1}^{n}\text{log}\left(1-y_{i}^{-2\gamma}\right)\right]}\sim\text{Gamma}\left(n,-\sum_{i=1}^{n}\text{log}\left(1-y_{i}^{-2\gamma}\right)\right)\label{eq:15}
\end{align}
and
\begin{align}
p\left(\gamma\left|\alpha,\boldsymbol{\bm{y}}\right.\right) & \propto\gamma^{n\alpha-1}e^{-2\gamma\sum_{i=1}^{n}\text{log}\left(y_{i}\right)}\sim\text{Gamma}\left(n\alpha,2\sum_{i=1}^{n}\text{log}\left(y_{i}\right)\right).\label{eq:16}
\end{align}
It now follows that the EVI of the $\text{\text{TLPa}}$(conditional
on $\alpha$ and $\boldsymbol{\bm{y}}$) is inverse gamma, $\text{EVI}_{\text{TLPa}}\left|\alpha,\boldsymbol{\bm{y}}\right.\sim\text{InvGamma}\left(n\alpha,\left(\sum_{i=1}^{n}\text{log}\left(y_{i}\right)\right)^{-1}\right)$,
where the second parameter is the rate parameter. The derivation of
(\ref{eq:15}) is trivial. The approximation of (\ref{eq:16}) is
given in the Appendix. Equations (\ref{eq:12}), (\ref{eq:13}) and
(\ref{eq:14}) will be used in the simulation study of Section \ref{sec:Simulation-study}.

\section{Simulation study \label{sec:Simulation-study}}

In this simulation study observations are simulated from heavy tailed
distributions such as the Fr\'echet and the Burr distributions where
$\gamma>0$. At each threshold level (from the smallest sorted observation
to the second largest sorted observation) the parameters of the two
models (SP and TLPa) are estimated. In the case of the SP, $\hat{\gamma}$
is estimated as the mean of the gamma distribution in (\ref{eq:12}).
A Gibbs sampler is considered for the TLPa. A starting value for $\gamma$
is chosen. For the chosen $\gamma$, an $\hat{\alpha}^{\ast}$ is
drawn from $\eqref{eq:15}$. Given $\hat{\alpha}^{\ast}$, a new $\hat{\gamma}^{\ast}$
is drawn from the joint posterior given in (\ref{eq:14}). The process
is repeated until 2000 pairs of $\left(\hat{\alpha}^{\ast},\hat{\gamma}^{\ast}\right)$
are simulated. The estimates of $\alpha$ and $\gamma$ are then taken
as the respective means of the 2000 simulated values. Recall that
as $n\rightarrow\infty$ the TLPa becomes the SP, with $\alpha=1$.
This implies that the choice of threshold can be simplified by choosing
the threshold as that sorted observation where $\alpha$ is closest
to one. This will be demonstrated in the following case studies.

\subsection*{Case 1}

\begin{figure}[H]
\begin{centering}
\includegraphics[width=0.8\textwidth]{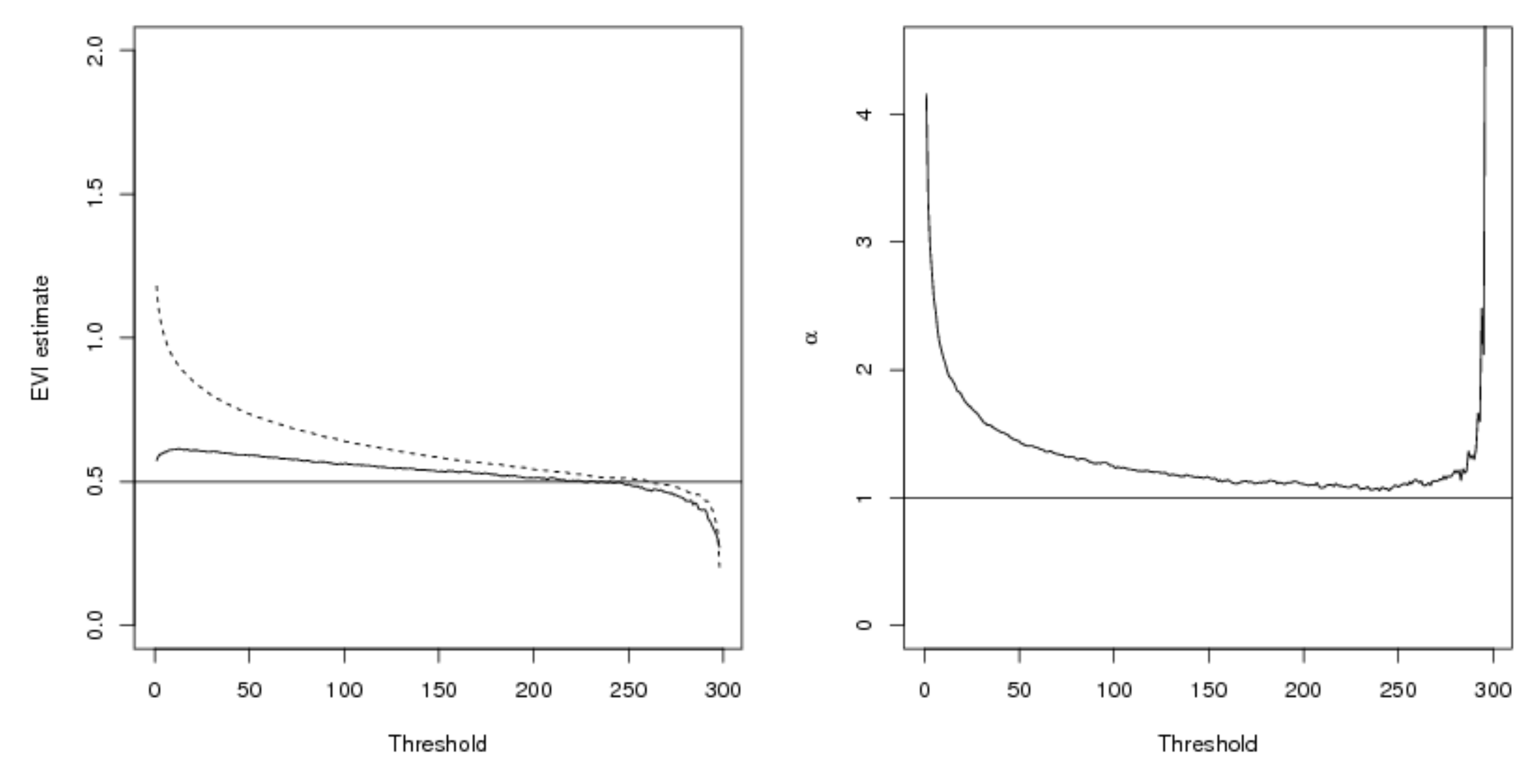}
\par\end{centering}
\caption{Left side: EVI estimates of the SP (dashed line) and TLPa (solid line).
Right side: $\alpha$ estimate to TLPa. }
\label{Case1}
\end{figure}

300 observations are simulated from a Fr\'echet distribution with $F\left(x\right)=e^{-x^{-\gamma}},\;x>0,\;\gamma=2$
where the $\text{EVI}=\nicefrac{1}{\gamma}=\nicefrac{1}{2}.$ The
simulation process is repeated 1000 times. Figure \ref{Case1} shows
the EVI estimates (of the SP (dashed line) and TLPa (solid line))
on the left side and the estimate of $\alpha$ in the TLPa model on
the right side. The parameter estimates were obtained as the average
over the 2000 posterior simulations as well as the average over the
1000 repetitions. Figure (\ref{Case1}) (right side) shows that $\alpha$
moves towards 1 as the threshold increases, $\alpha$ is the closest
to 1 for threshold values from the $220^{\text{th}}$ to $280^{\text{th}}$
sorted observations. This corresponds with the threshold values where
the EVI estimate (Figure \ref{Case1} left) reaches the true EVI estimate
of 0.5. 

\subsection*{Case 2}

\begin{figure}[H]
\begin{centering}
\includegraphics[width=0.8\textwidth]{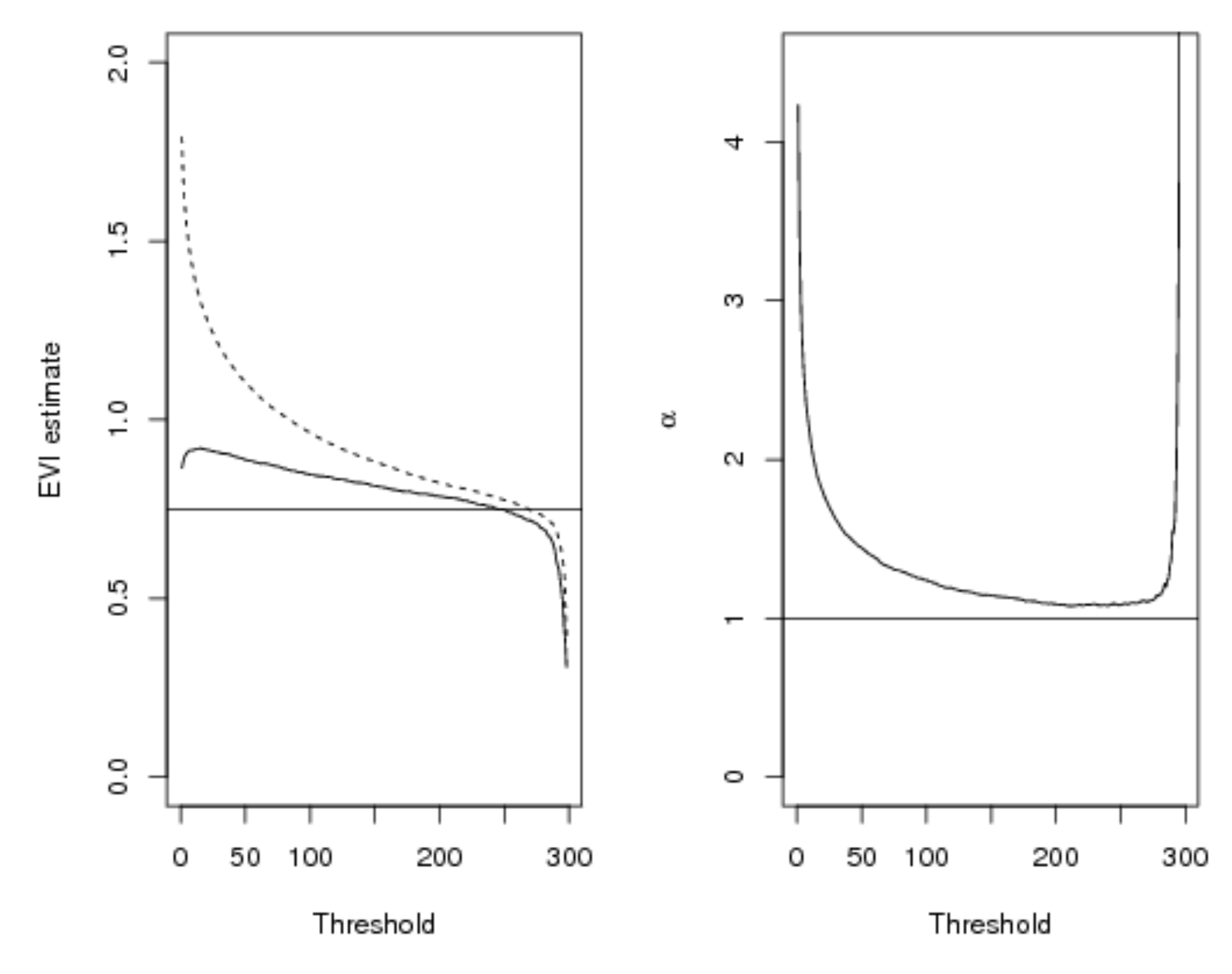}
\par\end{centering}
\caption{Left side: EVI estimates of the SP (dashed line) and TLPa (solid line).
Right side: $\alpha$ estimate to TLPa. }
\label{Case2}
\end{figure}

300 observations are simulated from a Fr\'echet distribution with $\gamma=1.33$
and $\text{EVI}=\nicefrac{1}{\gamma}=0.75.$ The outcomes are similar
to the outcomes from Figure \ref{Case1}. Figure \ref{Case2} shows
the EVI estimates (of the SP and TLPa) and the estimate of $\alpha$
in the TLPa model. $\alpha$ moves towards 1 around the $220^{\text{th}}$
to $260^{\text{th}}$ sorted observations. This corresponds to where
the $EVI$ estimate (Figure \ref{Case2} left) reaches the true EVI
value of 0.75.

\subsection*{Case 3}

\begin{figure}[H]
\begin{centering}
\includegraphics[width=0.8\textwidth]{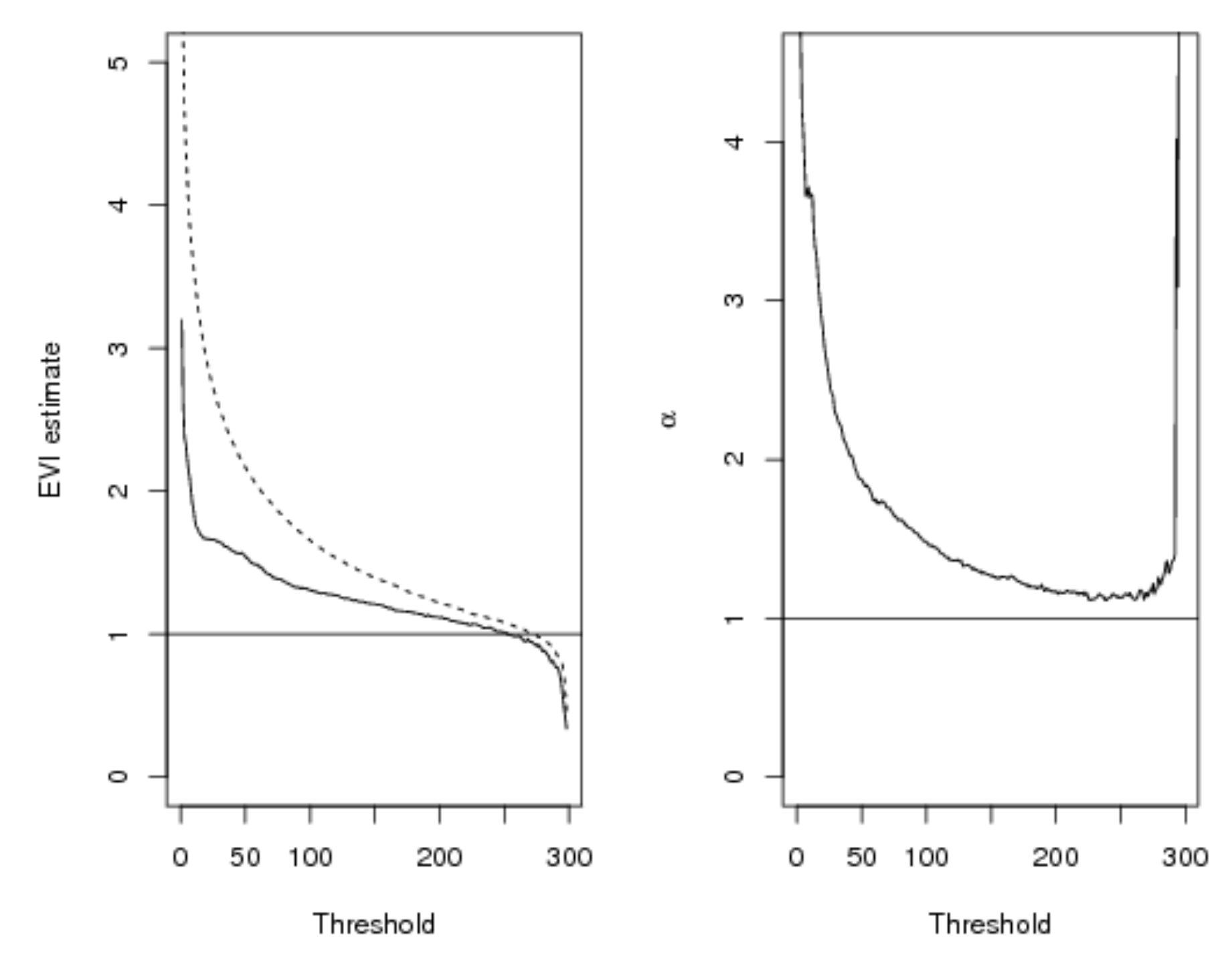}
\par\end{centering}
\caption{Left side: EVI estimates of the SP (dashed line) and TLPa (solid line).
Right side: $\alpha$ estimate to TLPa. }
\label{Case3}
\end{figure}

300 observations are simulated from the Burr Type XII distribution
with $F\left(x\right)=1-\left(\frac{\eta}{\eta+x^{\tau}}\right)^{\lambda},\;x,\eta,\tau,\lambda>0$
with $EVI=\nicefrac{1}{\lambda\tau}$. Let $\lambda=1$, $\tau=1$
and $\eta=1$, therefore the $\text{EVI}=1$. The simulation process
is again repeated 1000 times. Figure \ref{Case3} shows the EVI estimates
and the estimate of $\alpha$ in the TLPa model. The parameter estimates
were obtained as the average over the 2000 posterior simulations as
well as the average over the 1000 repetitions. Figure \ref{Case3}
shows that $\alpha$ moves towards 1 as the threshold increases and
is the closest to 1 for threshold values from the $240^{\text{th}}$
to the $280^{\text{th}}$ sorted observation. This corresponds with
the threshold values where the EVI estimate reaches the true $EVI$
estimate of 1. 

It can be seen from the above cases that the EVI estimates of the
TLPa is more stable, and less sensitive to the threshold choice, than
the EVI estimate of the SP. This is expected since the variance of
$\gamma_{\text{TLPa}}\left|\alpha,\boldsymbol{y}\right.,$ decreases
as the value of $\alpha$ increases, see (\ref{eq:12}) and (\ref{eq:16}).

\section{Real data example \label{sec:Real-data-example}}

\begin{figure}[H]
\begin{centering}
\includegraphics{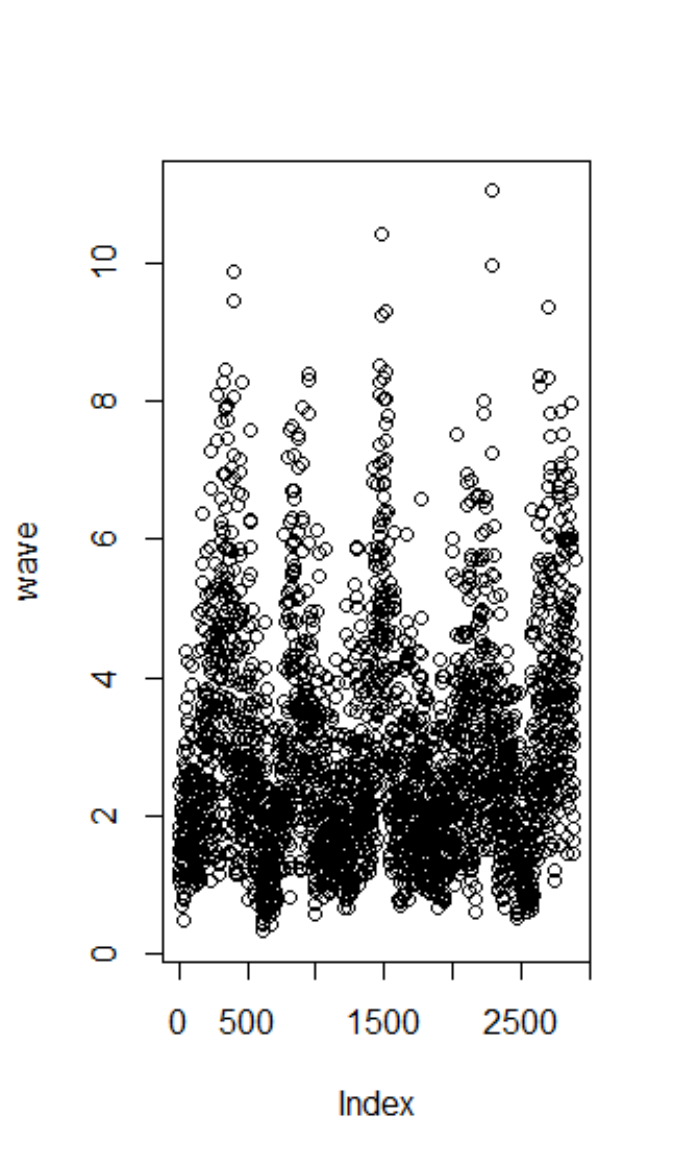}
\par\end{centering}
\caption{Wave height data observations.}
\label{AppFig1}
\end{figure}

\begin{figure}[H]
\begin{centering}
\includegraphics[width=0.8\textwidth]{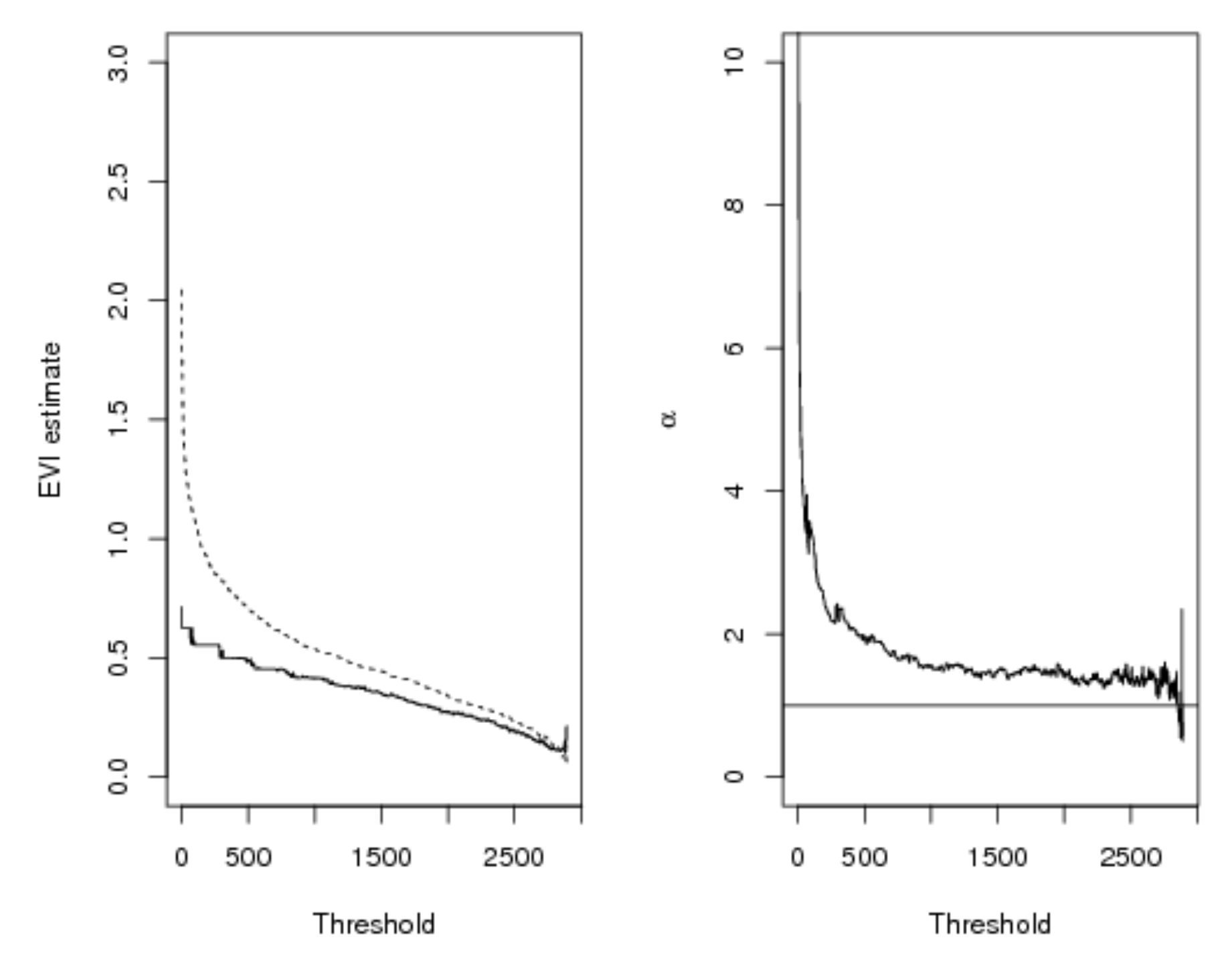}
\par\end{centering}
\caption{EVI estimates (top) and the estimate of $\alpha$ (bottom) in the
TLPa model.}
\label{AppFig2}
\end{figure}

In this section the wave height data from Newlyn, Cornwall, is considered
as a real data example. A POT approach is used to model this data
set and according the the mean residual plot a marginal threshold
is chosen at 6.1 meters, see Coles (2001). The data observations with
$n=2894$ is given in Figure \ref{AppFig1}. Extreme observations
can easily be observed from the figure. Figure \ref{AppFig2} shows
the EVI estimates of the two models (left) and the estimate of $\alpha$
(right) for the TLPa model. Figure \ref{AppFig2} shows that $\alpha$
moves towards 1 around the $2800^{\text{th}}$ sorted observation.
This corresponds to a wave height of 6.67 meters which is close to
the 6.1 meters threshold chosen by Coles (2001). The data set can
be obtained from the \texttt{ismev} package in \textsf{\textbf{R}}.

\begin{figure}[H]
\begin{centering}
\includegraphics[width=0.8\textwidth]{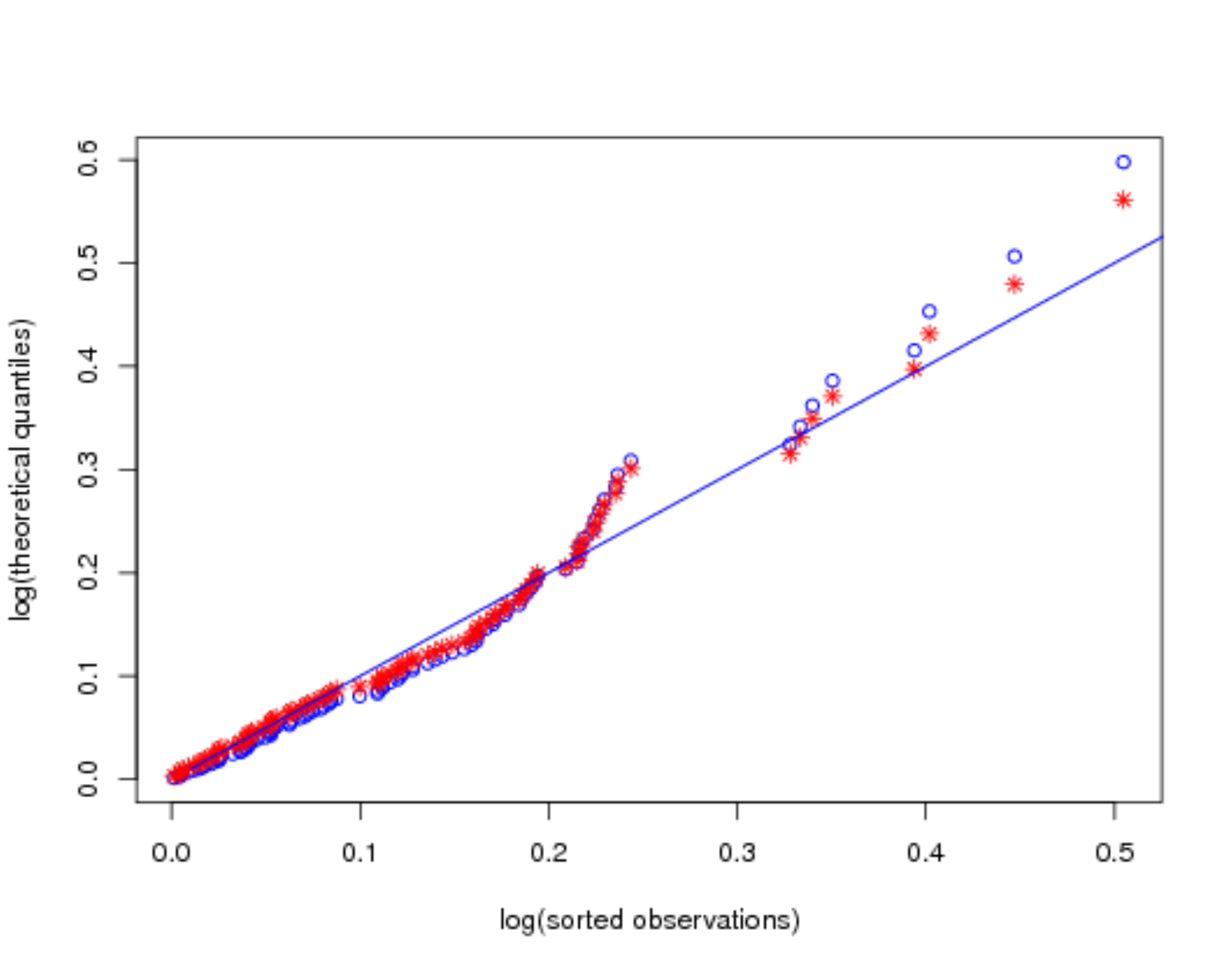}
\par\end{centering}
\caption{The log of the SP quantiles (in blue $\circ$) and the log of the
TLPa quantiles (in red $*$) against the log of the sorted observations.}
\label{AppFig3}
\end{figure}

The threshold was chosen as the $2800^{\text{th}}$ sorted observation
and the quantiles (above the chosen threshold) for these two models
were calculated by assuming the estimated parameter values at this
threshold. As shown in Figure \ref{AppFig3} the log quantile-quantile
plot of both distributions indicates a good fit. This is expected
since the threshold of 2800 seems to be optimal from Figure \ref{AppFig2}. 

\section{Procedure to choose a threshold \label{sec:Procedure-to-choose}}

Since the main focus of this paper is on threshold choice and not
bias reduction this section introduces a practical and modest method
to choose a threshold. The conditional posterior of $\alpha$ from
(\ref{eq:15}) is used. The advantage of this method is that the threshold
can be chosen without visually inspecting a graph. Since $E\left(\alpha\left|\gamma,\boldsymbol{\bm{y}}\right.\right)=n\left(-\sum_{i=1}^{n}\text{log}\left(1-y_{i}^{-2\gamma}\right)\right)^{-1}$
and using the theory that as $n\rightarrow\infty$, $\alpha\rightarrow1$,
the method proposes to find the combination of $\gamma^{\#}$ and
$u^{\#}$ (from a large spectrum of $\gamma$ and $u$ values) for
which the Bayes estimate under squared error loss, $\left[E\left(\alpha\left|\gamma,\boldsymbol{\bm{y}}\right.\right)-1\right]^{2}$,
is a minimum. Considering the above method two data sets are simulated
for illustration purposes

i. $n=300$ observations from Fr\'echet $\left(\gamma=2\right)$.

ii. $n=300$ observations from Burr XII $\left(\lambda=1,\tau=1,\eta=1\right)$. 

\begin{figure}[H]
\begin{centering}
\includegraphics[width=0.8\textwidth]{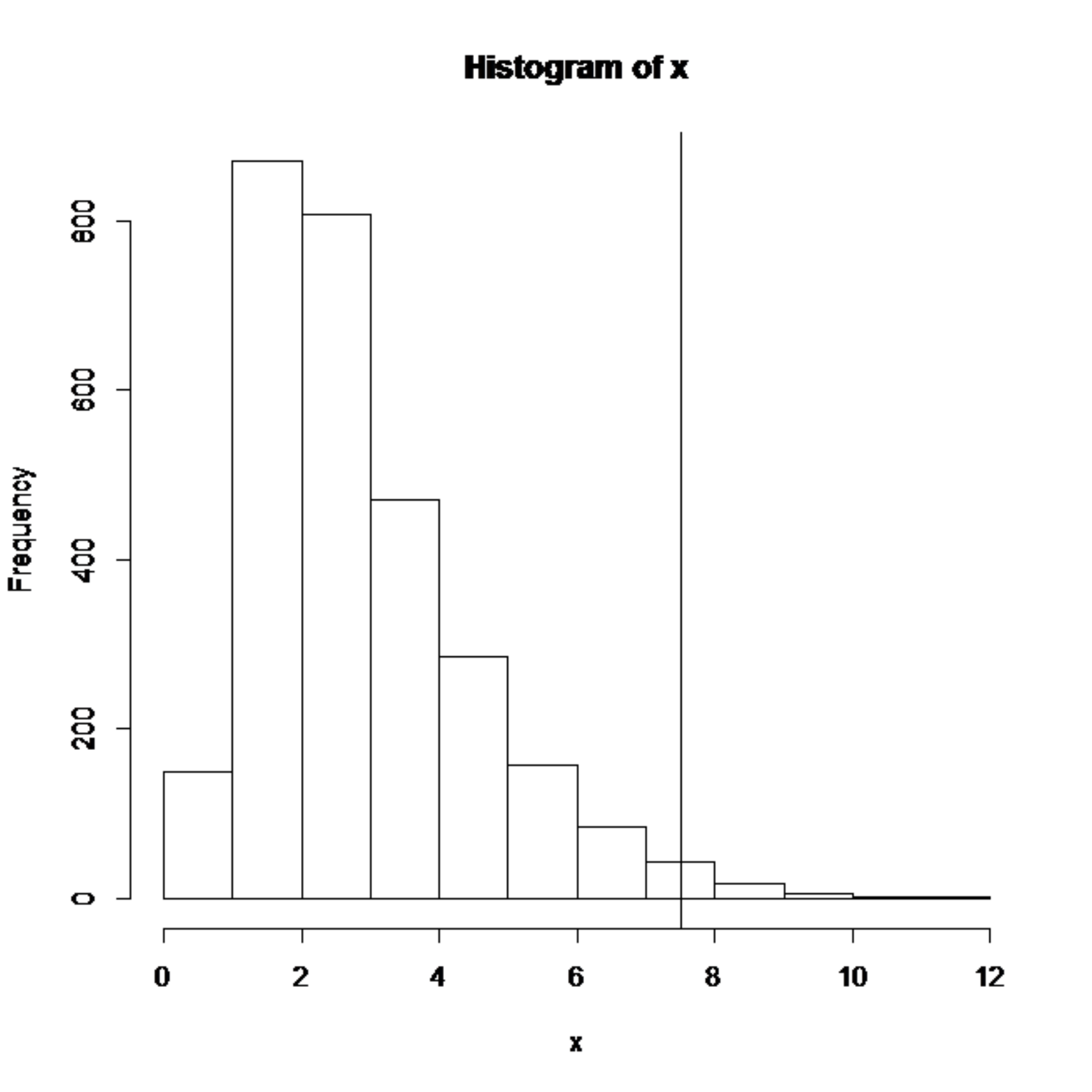}
\par\end{centering}
\caption{Histogram with the chosen threshold for the wave height data.}
\label{Thresh1}
\end{figure}

For data set (i) the $\text{EVI}_{\text{TLPa}}=\nicefrac{1}{\left(2\gamma^{\#}\right)}$
is chosen as 0.5383 and $u^{\#}$ is chosen as the $174^{\text{th}}$
sorted observation. For data set (ii) the $\text{EVI}_{\text{TLPa}}=\nicefrac{1}{\left(2\gamma^{\#}\right)}$
is chosen as 1.1230 and $u^{\#}$ is chosen as the $196^{\text{th}}$
sorted observation. For the real data example the $\text{EVI}_{\text{TLPa}}=\nicefrac{1}{\left(2\gamma^{\#}\right)}$
is chosen as 0.1158 and $u^{\#}$ is chosen as the $2850^{\text{th}}$
sorted observation ($u=7.52$). These chosen thresholds, together
with the EVI estimates seem to be in line with the results obtained
from the simulation studies (Figures \ref{Case1} and \ref{Case3})
and the wave height application (Figure \ref{AppFig2}). The histogram
with the chosen threshold (for the wave height date) is shown in Figure
\ref{Thresh1}. 

Another two data sets are simulated to test the appropriateness of
the above method. 

iii. $n_{1}=500$ observations from $\text{Normal}(\mu=5,\sigma^{2}=1)$
and $n_{2}=100$ observations from SP$(\gamma=5)$ and $u$ is chosen
as the maximum Normal observation. The two simulated data sets are
joined together as a single data set. The SP threshold is thus the
$500^{\text{th}}$ sorted observation. 

iv. $n_{1}=500$ observations from $\text{Normal}(\mu=10,\sigma^{2}=16)$
and $n_{2}=100$ observations from SP$(\gamma=2)$ and $u$ is chosen
again as the maximum Normal observation. The two simulated data sets
are joined together as a single data set. Again the SP threshold is
the $500^{\text{th}}$ sorted observation. 

\begin{figure}[H]
\begin{centering}
\includegraphics[width=0.8\textwidth]{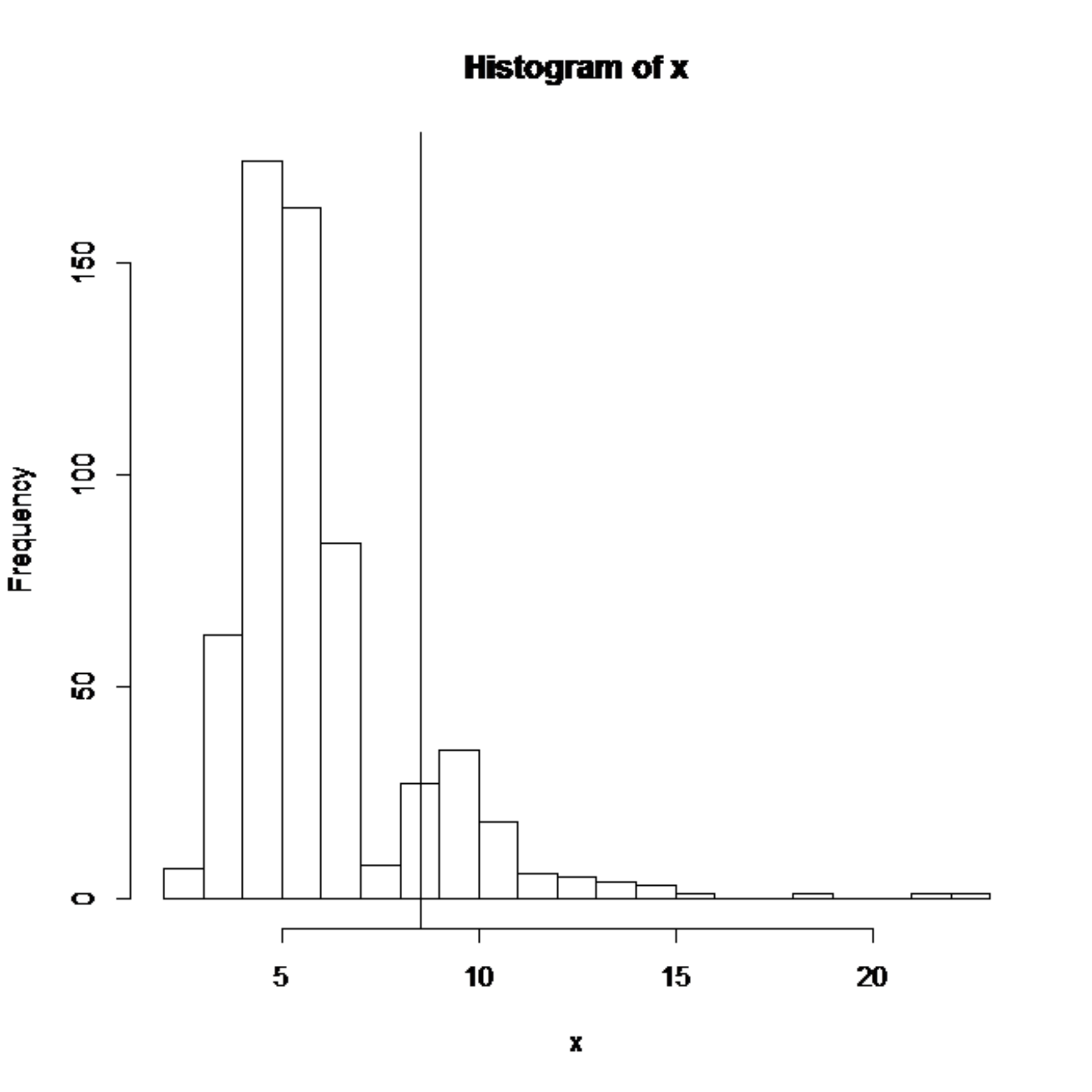}
\par\end{centering}
\caption{Histogram of data set (iii) with the chosen threshold indicated with
the vertical line.}
\label{Thresh2}
\end{figure}

Figure \ref{Thresh2} shows the histogram of data set (iii) where
the threshold is indicated with a vertical line. From Figure \ref{Thresh2}
it seems quite easy to visually distinguish between the two distributions.
The threshold method is applied to the simulated data set where the
threshold was chosen from the 50 percentile onward. An EVI-threshold
pair that minimizes the Bayes estimate under squared error loss is
selected. The process is repeated 1000 times. Table \ref{Table1}
shows the mean of the 1000 chosen thresholds as well as the mean of
the $\text{EVI}_{\text{TLPa}}$ estimates. The chosen threshold seems
to be rather close to the actual threshold and the EVI estimate is
close to the true EVI, $\nicefrac{1}{\gamma}=0.2$. 

\begin{figure}[H]
\begin{centering}
\includegraphics[width=0.8\textwidth]{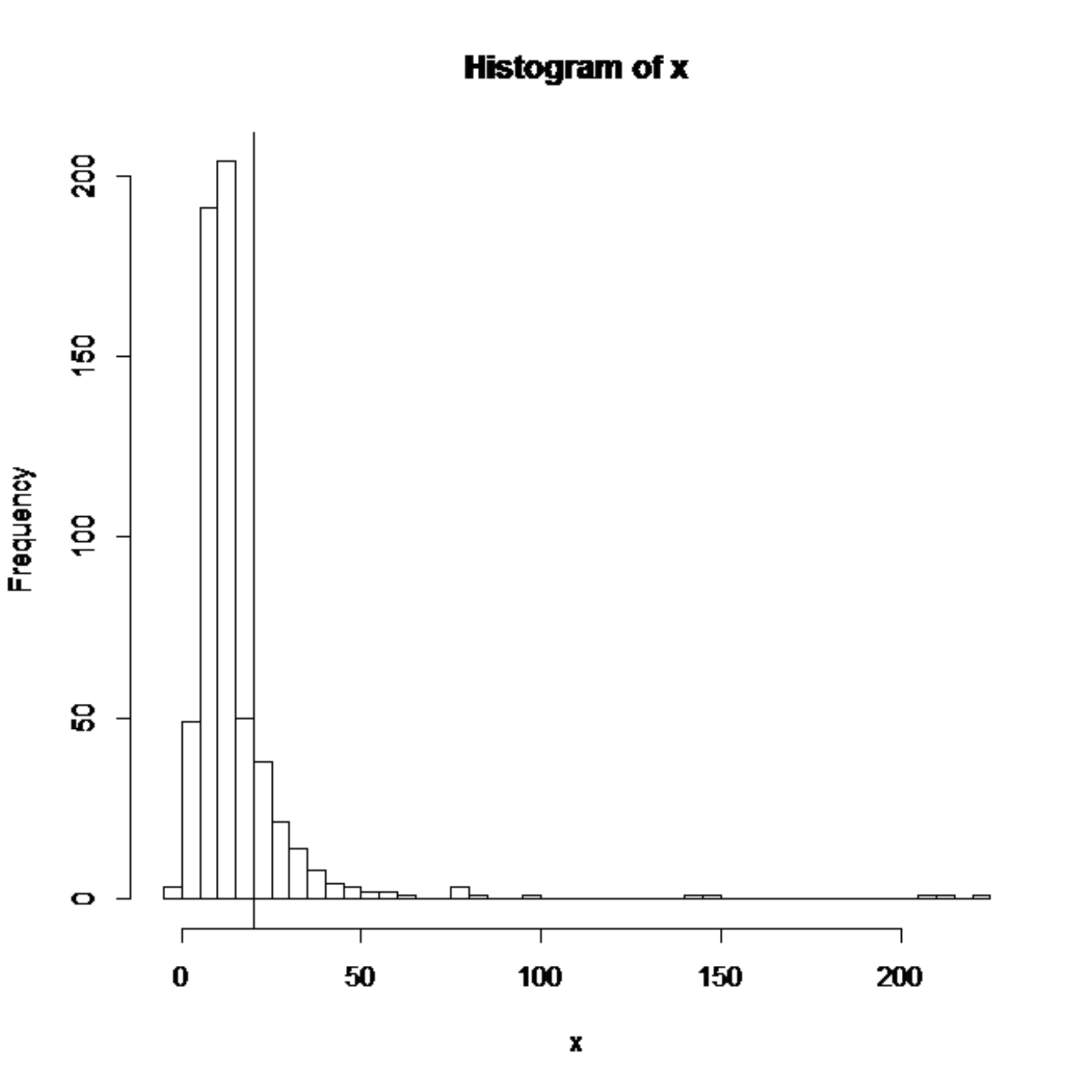}
\par\end{centering}
\caption{Histogram of data set (iv) with the chosen threshold indicated with
the vertical line.}
\label{Thresh3}
\end{figure}

Figure \ref{Thresh3} shows the histogram of data set (iv) where the
threshold is indicated with a vertical line. From Figure \ref{Thresh3}
it is not visually possible to distinguish between the two distributions,
giving the smooth transition from the Normal distribution to the SP
distribution. Table \ref{Table2} shows the mean of the 1000 chosen
thresholds as well as the mean of the $\text{EVI}_{\text{TLPa}}$
estimates. Since the transition between the two distribution are rather
smooth the chosen threshold is not as close to the actual threshold
but still reasonable and the $\text{EVI}_{\text{TLPa}}$ estimate
is still rather close to the true EVI, $\nicefrac{1}{\gamma}=0.5$. 

\begin{table}[H]
\caption{Threshold estimate and $\text{EVI}_{\text{TLPa}}$ estimate.}
\label{Table1}
\begin{centering}
\begin{tabular}{|c|c|}
\hline 
\textbf{Threshold estimate} & \textbf{$EVI$ estimate}\tabularnewline
\hline 
457.339 & 0.2337\tabularnewline
\hline 
\end{tabular}
\par\end{centering}
\end{table}

\begin{table}[H]
\caption{Threshold estimate and $\text{EVI}_{\text{TLPa}}$ estimate.}
\label{Table2}
\centering{}%
\begin{tabular}{|c|c|}
\hline 
\textbf{Threshold estimate} & \textbf{$EVI$ estimate}\tabularnewline
\hline 
442.786 & 0.4980865\tabularnewline
\hline 
\end{tabular}
\end{table}

\section{Conclusion}

In this study we have shown that the TLPa is a suitable model to use
when modelling extreme events above a threshold when $\gamma>0$.
The TLPa has proven to be less sensitive to the correct threshold
choice. The focus of the study was to show that the $\alpha$ parameter
of the TLPa can successfully contribute to choosing an appropriate
threshold without using any visual technique. 

A Bayesian approach was considered and the conditional posterior distributions
of the two parameters of the TLPa was derived. These posteriors were
valuable in the simulation studies. A method for choosing an optimum
threshold was introduced. This method involves finding a combination
of $\text{EVI}_{\text{TLPa}}$ and $u$ (threshold) values that minimizes
the Bayes estimate under squared error loss. The illustrations in
Section \ref{sec:Procedure-to-choose} shows that the method is easy
to use with sufficient results.

\section*{References}
\begin{list}{}{}
\item Beirlant, J., Dierckx, G., Goegebeur, Y., \& Matthys, G. (1999). Tail index estimation and an exponential
regression model. \emph{Extremes}, 2(2), 177-200.
\item Beirlant, J., Goegebeur, Y., Segers, J., \& Teugels, J. (2004). \emph{Statistics of Extremes: Theory and Applications}.
Chichester: John Wiley and Sons.
\item Beirlant, J., Joossens, E., \& Segers, J. (2009). Second-order refined peaks-over-threshold modelling for
heavy-tailed distribution. \emph{Journal of Statistical Planning and Inference}, 139, 2800-2815.
\item Beirlant, J., Maribe, G., \& Verster, A. (2019). Using shrinkage estimators to reduce bias and mse in
estimation of heavy tails. \emph{REVSTAT-Statistical Journal}, 17(1), 91-108.
\item Caeiro, F. \& Gomes, M. I. (2011). Asymptotic comparison at optimal levels of reduced-bias extreme
value index estimators. \emph{Statistica Neerlandica}, 65(4), 462-488.
\item Coles, S. G. (2001). \emph{An introduction to statistical modeling of extreme values}. London: Springer.
\item de Haan, L. \& Ferreira, A. (2005). \emph{Extreme Value Theory: An Introduction}. New York: Springer
\item Feuerverger, A. \& Hall, P. (1999). Estimating a tail exponent by modelling departure from a Pareto
distribution. \emph{The Annals of Statistics}, 27(2), 760-781.
\item Gomes, M. I., Martins, M. J., \& Neves, M. (2000). Alternatives to a semiparametric estimator of parameters
of rare events - the jackknife methodology. \emph{Extremes}, 3(3), 207-229.
\item Rezaei, S., Sadr, B. B., Alizadeh, M., \& Nadarajah, S. (2017). Topp-Leone generated family of distributions:
Properties and applications. \emph{Communications in Statistics - Theory and Methods}, 46(6), 2893-2909.
\item Topp, C. W. \& Leone, F. C. (1955). A family of J-shaped frequency functions. \emph{Journal of the American
Statistical Association}, 50, 209-219.
\end{list}{}{}

\section*{Appendix}

Approximating the conditional posterior distribution of $\gamma$
given $\alpha$ and $\boldsymbol{y}$.

From the joint posterior in (\ref{eq:14}) the conditional posterior
is 
\begin{align*}
p\left(\gamma\left|\alpha,\boldsymbol{y}\right.\right) & \propto\gamma^{n-1}e^{-2\gamma\sum_{i=1}^{n}\text{log}\left(y_{i}\right)}e^{\left(\alpha-1\right)\sum_{i=1}^{n}\text{log}\left(1-y_{i}^{-2\gamma}\right)}.
\end{align*}
By using the Taylor series expansion, $e^{\left(\alpha-1\right)\sum_{i=1}^{n}\text{log}\left(1-y_{i}^{-2\gamma}\right)}$
can be approximated as $\prod_{i=1}^{n}\left(2\gamma\text{log}y_{i}\right)^{\alpha-1}$. 

Therefore
\begin{align*}
p\left(\gamma\left|\alpha,\boldsymbol{y}\right.\right) & \propto\gamma^{n-1}e^{-2\gamma\sum_{i=1}^{n}\text{log}\left(y_{i}\right)}\prod_{i=1}^{n}\left(2\gamma\text{log}y_{i}\right)^{\alpha-1}\\
 & \propto\gamma^{n\alpha-1}e^{-2\gamma\sum_{i=1}^{n}\text{log}\left(y_{i}\right)}.
\end{align*}

\end{document}